\documentclass[acmsmall]{acmart}

\setcopyright{rightsretained}
\copyrightyear{2021}
\acmYear{2021}
\acmConference[HATRA '21]{Human Aspects of Types and Reasoning Assistants}{October 2021}{Chicago, IL}
\acmBooktitle{HATRA '21: Human Aspects of Types and Reasoning Assistants}
\acmDOI{}
\acmISBN{}
\expandafter\def\csname @copyrightpermission\endcsname{\raisebox{-1ex}{\includegraphics[height=3.5ex]{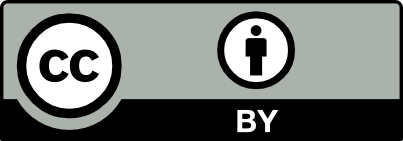}} This work is licensed under a Creative Commons Attribution 4.0 International License.}
\expandafter\def\csname @copyrightowner\endcsname{Roblox.}

\newcommand{\squnder}[1]{\color{red}\underline{{\color{black}#1}}\color{black}}

\newcommand{\erase}{\mathrm{erase}}
\newcommand{\evCtx}{\mathcal{E}}
\newcommand{\NIL}{\mathsf{nil}}
\newcommand{\ANY}{\mathsf{any}}
\newcommand{\TRUE}{\mathsf{true}}

\newcommand{\NUMBER}{\mathsf{number}}
\newcommand{\STRING}{\mathsf{string}}
\newcommand{\ERROR}{\mathsf{error}}
\newcommand{\IF}{\mathsf{if}\,}
\newcommand{\LOCAL}{\mathsf{local}\,}
\newcommand{\THEN}{\,\mathsf{then}\,}

\newcommand{\END}{\,\mathsf{end}}

\newcommand{\FIND}{\mathsf{find}}
\newcommand{\PRINT}{\mathsf{print}}
\newcommand{\strlit}[1]{\mbox{``#1''}}

\begin{document}

\title{Position Paper: Goals of the Luau Type System}

\author{Lily Brown}
\author{Andy Friesen}
\author{Alan Jeffrey}
\affiliation{
  \institution{Roblox}
  \city{San Mateo}
  \state{CA}
  \country{USA}
}

\begin{abstract}
  Luau is the scripting language that powers user-generated experiences on the
  Roblox platform. It is a statically-typed language, based on the
  dynamically-typed Lua language, with type inference. These types are used for providing
  editor assistance in Roblox Studio, the IDE for authoring Roblox experiences.
  Due to Roblox's uniquely heterogeneous developer community, Luau must operate
  in a somewhat different fashion than a traditional statically-typed language.
  In this paper, we describe some of the goals of the Luau type system,
  focusing on where the goals differ from those of other type systems.
\end{abstract}

\maketitle

\section{Introduction}

The Roblox platform allows anyone to create shared,
immersive, 3D experiences.  As of July 2021, there are
approximately 20~million experiences available on Roblox, created
by 8~million developers~\cite{Roblox}.  Roblox creators are often young: there are
over 200~Roblox kids' coding camps in 65~countries
listed by the company as education resources~\cite{AllEducators}.
The Luau programming language~\cite{Luau} is the scripting language
used by creators of Roblox experiences. Luau is derived from the Lua
programming language~\cite{Lua}, with additional capabilities,
including a type inference engine.

This paper will discuss some of the goals of the Luau type system, such
as supporting goal-driven learning, non-strict typing semantics, and
mixing strict and non-strict types.  Particular focus is placed on how
these goals differ from traditional type systems' goals.

\section{Needs of the Roblox platform}
\subsection{Heterogeneous developer community}

Need: \emph{a language that is powerful enough to support professional users, yet accessible to beginners}

Quoting a Roblox 2020 report \cite{RobloxDevelopers}:
\begin{itemize}
\item \emph{Adopt Me!} now has over 10 billion plays and surpassed 1.6 million concurrent users earlier this year.
\item \emph{Piggy}, launched in January 2020, has close to 5 billion visits in just over six months.
\item There are now 345,000 developers on the platform who are monetizing their games.
\end{itemize}
This demonstrates the heterogeneity of the Roblox developer community:
developers of experiences with billions of plays are on the same
platform as children first learning to code. Both of these groups are important to
support: the professional development studios bring high-quality experiences to the
platform, and the beginning creators contribute to the energetic creative community,
forming the next generation of developers.

\subsection{Goal-driven learning}

Need: \emph{organic learning for achieving specific goals}

All developers are goal-driven, but this is especially true for
learners. A learner will download Roblox Studio
(the creation environment for the Roblox platform) with an
experience in mind, such as designing an obstacle course
to play in with their friends.

\begin{figure}
\includegraphics[width=0.48\textwidth]{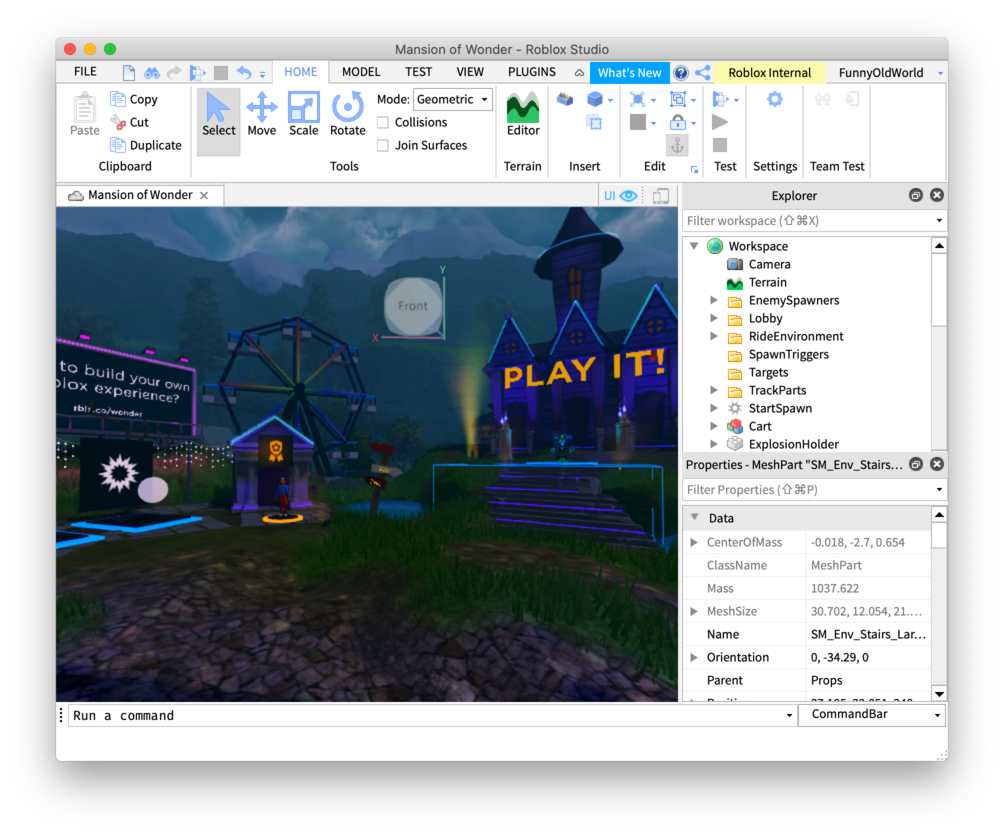}
\includegraphics[width=0.48\textwidth]{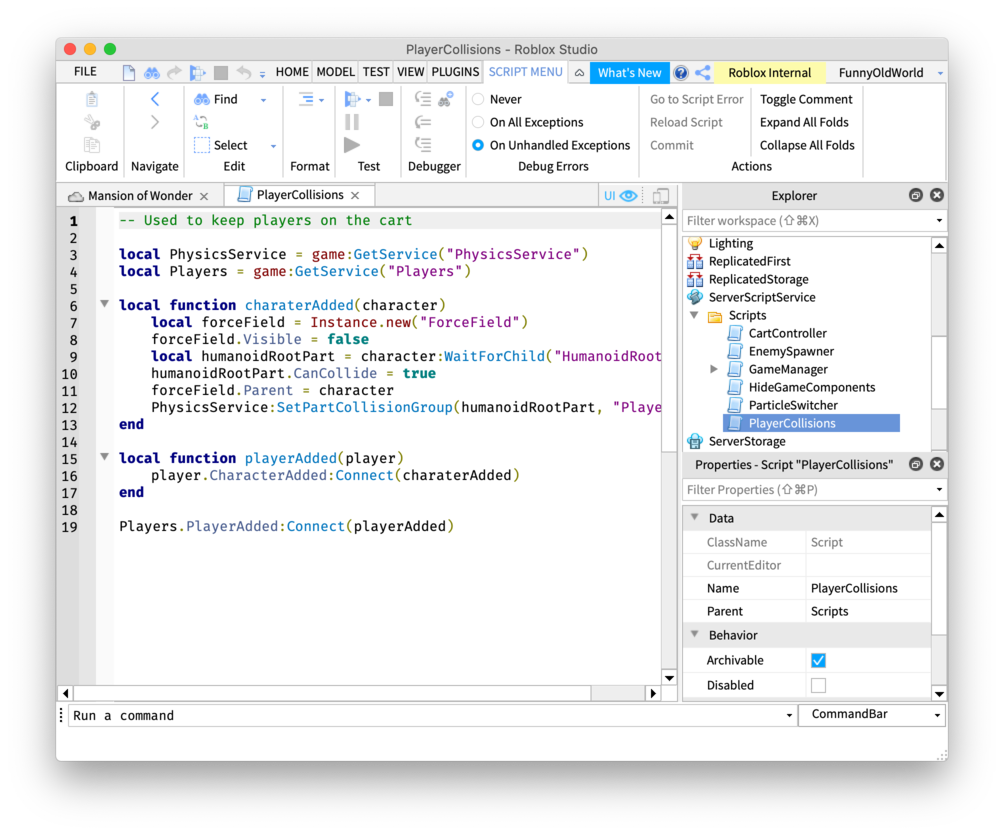}
\caption{Roblox Studio's 3D environment editor (a), and script editor (b)}
\label{fig:studio}
\end{figure}

The user experience of developing a Roblox experience is primarily a
3D interactive one, seen in Fig.~\ref{fig:studio}(a). The user designs
and deploys 3D assets such as terrain, parts and joints, providing
them with physics attributes such as mass and orientation. The user
can interact with the experience in Studio, and deploy it to a Roblox
server so anyone with the Roblox app can play it. Physics, rendering
and multiplayer are all immediately accessible to creators.

At some point during experience design, the experience creator has a need
that can't be met by the game engine alone, such as ``the stairs should
light up when a player walks on them'' or ``a firework is set off
every few seconds''. At this point, they will discover the script
editor, seen in Fig.~\ref{fig:studio}(b).

This onboarding experience is different from many initial exposures to
programming, in that by the time the user first opens the script
editor, they have already built much of their creation, and have a
very specific concrete aim. As such, Luau must allow users to perform a
specific task with as much help as possible from tools.

A common workflow for getting started is to Google the task, then
cut-and-paste the resulting code, adapting it as needed.  Since this
is so common, backward compatibility of Luau with existing code is
important, even for learners who do not have an existing code base to
maintain.

Type-driven tools are useful to all creators, in as much as they help
them achieve their current goals. For example type-driven
autocomplete, or type-driven API documentation, are of immediate
benefit. Traditional typechecking can be useful, for example for
catching spelling mistakes, but for most goal-driven developers, the
type system should help or get out of the way.

\subsection{Type-driven development}

Need: \emph{a language that supports large-scale codebases and defect detection}

Professional development studios are also goal-directed (though the
goals may be more abstract, such as ``decrease user churn'' or
``improve frame rate'') but have additional needs:
\begin{itemize}

\item \emph{Code planning}:
  code spends much of its time in an incomplete state, with holes
  that will be filled in later.

\item \emph{Code refactoring}:
  code evolves over time, and it is easy for changes to
  break previously-held invariants.

\item \emph{Defect detection}:
  code has errors, and detecting these at runtime (for example by crash telemetry)
  can be expensive and recovery can be time-consuming.
  
\end{itemize}
Detecting defects ahead-of-time is a traditional goal of type systems,
resulting in an array of techniques for establishing safety results,
surveyed for example in~\cite{TAPL}. Supporting code planning and
refactoring are some of the goals of \emph{type-driven
development}~\cite{TDDIdris} under the slogan ``type, define,
refine''.  A common use of type-driven development is renaming a
property, which is achieved by changing the name in one place,
and then fixing the resulting type errors---once the type system stops
reporting errors, the refactoring is complete.

To help support the transition from novice to experienced developer,
types are introduced gradually, through API documentation and type discovery.
Type inference provides many of the benefits of type-driven development
even to creators who are not explicitly providing types.

\section{Goals of the type system}
\subsection{Infallible types}

Goal: \emph{provide type information even for ill-typed or syntactically invalid programs.}

Programs spend much of their time under development in an ill-typed or incomplete state, even if the
final artifact is well-typed. If tools such as autocomplete and API documentation are type-driven,
this means that tooling needs to rely on type information even for ill-typed
or syntactically invalid programs. An analogy is infallible parsers, which perform error recovery and
provide an AST for all input texts, even if they don't adhere to the parser's syntax.

Program analysis can still flag type errors, which may be presented
to the user with red squiggly underlining. Formalizing this, rather
than a judgment
$\Gamma\vdash M:T$, for an input term $M$, there is a judgment
$\Gamma \vdash M \Rightarrow N : T$ where $N$ is an output term
where some subterms are \emph{flagged} as having type errors, written $\squnder{N}$. Write $\erase(N)$
for the result of erasing flaggings: $\erase(\squnder{N}) = \erase(N)$.

For example, in Lua, the $\STRING.\FIND$ function expects two strings, and returns the
offsets for that string:
\[
  \STRING.\FIND(\strlit{hello}, \strlit{ell}) \rightarrow (2, 4)
\qquad
  \STRING.\FIND(\strlit{world}, \strlit{ell}) \rightarrow (\NIL, \NIL)
\]
and in Luau it has the type:
\[
  \STRING.\FIND : (\STRING, \STRING) \rightarrow (\NUMBER?, \NUMBER?)
\]
In a conventional type system, there is no judgment for ill-typed terms
such as $\STRING.\FIND(\strlit{hello}, 37)$ but in an infallible system we flag the error
and approximate the type, for example:
\[
  {} \vdash
  \STRING.\FIND(\strlit{hello}, 37)
  \Rightarrow
  \squnder{\STRING.\FIND(\strlit{hello}, 37)}
  :
  (\NUMBER?, \NUMBER?)
\]
The goal of infallible types is that every term has a typing judgment
given by flagging ill-typed subterms:
\begin{itemize}
\item \emph{Typability}: for every $M$ and $\Gamma$,
  there are $N$ and $T$ such that $\Gamma \vdash M \Rightarrow N : T$.
\item \emph{Erasure}: if $\Gamma \vdash M \Rightarrow N : T$
  then $\erase(M) = \erase(N)$ 
\end{itemize}
Some issues raised by infallible types:
\begin{itemize}
\item Which heuristics should be used to provide types for flagged programs? For example, could one
  use minimal edit distance to correct for spelling mistakes in field names?
\item How can we avoid cascading type errors, where a developer is
  faced with type errors that are artifacts of the heuristics, rather
  than genuine errors?
\item How can the goals of an infallible type system be formalized?
\end{itemize}
\emph{Related work}:
there is a large body of work on type error reporting
(see, for example, the survey in~\cite[Ch.~3]{TopQuality})
and on type-directed program repair
(see, for example, the survey in~\cite[Ch.~3]{RepairingTypeErrors}),
but less on type repair.
The closest work is Hazel's~\cite{Hazel} \emph{typed holes}
where $\squnder{N}$ is treated as a partially-filled hole in the program,
though in that work partially-filled holes are not erased at run-time.
Many compilers perform
error recovery during typechecking, but do not provide a semantics
for programs with type errors.

\subsection{Strict types}

Goal: \emph{no false negatives.}

For developers who are interested in defect detection, Luau provides a \emph{strict mode},
which acts much like a traditional, sound, type system. This has the goal of ``no false negatives''
where any possible run-time error is flagged. This is formalized using:
\begin{itemize}
\item \emph{Operational semantics}: a reduction judgment $M \rightarrow N$ on terms.
\item \emph{Values}: a subset of terms representing a successfully completed evaluation.
\end{itemize}
Error states at runtime are represented as stuck states (terms that are not
values but cannot reduce), and showing that no well-typed program is
stuck. This is not true if typing is infallible, but can fairly
straightforwardly be adapted. We extend the operational semantics to flagged terms,
where $M \rightarrow M'$ implies $\squnder{M} \rightarrow \squnder{M'}$, and
for any value $V$ we have $\squnder{V} \rightarrow V$, then show:
\begin{itemize}
\item \emph{Progress}: if ${} \vdash M \Rightarrow N : T$, then either $N \rightarrow N'$ or $N$ is a value or $N$ has a flagged subterm.
\item \emph{Preservation}: if ${} \vdash M \Rightarrow N : T$ and $N \rightarrow N'$ then  $M \rightarrow^*M'$ and ${} \vdash M' \Rightarrow N' : T$.
\end{itemize}
For example in typechecking the program:
\[
  \LOCAL (i,j) = \STRING.\FIND(x, y);
  \IF i \THEN \PRINT(j-i) \END
\]
the interesting case is $i-j$ in a context where  $i$ has type
$\NUMBER$ (since it is guarded by the $\IF$) but $j$ has type
$\NUMBER?$. Since subtraction has type $(\NUMBER, \NUMBER) \rightarrow \NUMBER$,
this is a type error, so the relevant typing judgment is:
\[\begin{array}{r@{}l}
  x: \STRING, y: \STRING \vdash {}&
  (\LOCAL (i,j) = \STRING.\FIND(x, y);
  \IF i \THEN \PRINT(j-i) \END) \\
  \Rightarrow {}&
  (\LOCAL (i,j) = \STRING.\FIND(x, y);
  \IF i \THEN \PRINT(\squnder{j-i}) \END)
\end{array}\]
Some issues raised by soundness for infallible types:
\begin{itemize}
\item How should the judgments and their metatheory be set up?
\item How should type inference and generic functions be handled?
\item Is the operational semantics of flagged values
  ($\squnder{V} \rightarrow V$) the right one?
\end{itemize}
\emph{Related work}: gradual typing and blame analysis, e.g.~\cite{GradualTyping,WellTyped,Contracts}.
The main difference between this approach and that of migratory typing~\cite{MigratoryTyping}
is that (due to backward compatibility with existing Lua) we cannot introduce
extra code during migration.

\subsection{Nonstrict types}

Goal: \emph{no false positives.}

For developers who are not interested in defect detection, type-driven
tools and techniques such as autocomplete, API documentation
and type-driven refactoring are still useful.
For such developers, Luau provides a
\emph{nonstrict mode}, which we hope will eventually be useful for all
developers.  This non-strict typing mode is particularly useful when
adopting Luau types in pre-existing code that was not authored with
the type system in mind.  Non-strict mode does \emph{not} aim for
soundness, but instead has the goal of ``no false positives``, in the
sense that any flagged code is guaranteed to produce a runtime error
when executed.

Our previous example was, in fact, a false positive since a programmer
can make use of the fact that $\STRING.\FIND(x, y)$ is either $\NIL$
in both results or neither, so if $i$ is non-$\NIL$ then so is $j$.
This is discussed in the English-language documentation but not reflected
in the type. So flagging $(i - j)$ is a false positive.

On the face of it, detecting all errors without false positives is undecidable, since a program such as
$(\IF f() \THEN \ERROR \END)$ will produce a runtime error when $f()$ is
$\TRUE$. Instead we can aim for a weaker property: that all flagged code
is either dead code or will produce an error. Either of these is a
defect, so deserves flagging, even if the tool does not know
which reason applies.

We can formalize this by defining an \emph{evaluation context}
$\evCtx[\bullet]$, and saying $M$ is \emph{incorrectly flagged}
if it is of the form $\evCtx[\squnder{V}]$. We can then define:
\begin{itemize}
\item \emph{Correct flagging}: if ${} \vdash M \Rightarrow N : T$
  then $N$ is correctly flagged.
\end{itemize}
Some issues raised by nonstrict types:
\begin{itemize}

\item Will nonstrict types result in errors being flagged in function call sites
  rather than definitions?

\item In Luau, ill-typed property update of most tables succeeds
  (the property is inserted if it did not exist), and so functions which
  update properties cannot be flagged. Can we still provide meaningful
  error messages in such cases?

\item Does nonstrict typing require whole program analysis,
  to find all the possible types a property might be updated with?

\item The natural formulation of function types in a nonstrict setting
  is that of~\cite{SuccessTyping}: if $f: T \rightarrow U$ and $f(V) \rightarrow^* W$
  then $V:T$ and $W:U$. This formulation is \emph{covariant} in $T$,
  not \emph{contravariant}; what impact does this have?
  
\end{itemize}
\emph{Related work}: success types~\cite{SuccessTyping} and incorrectness logic~\cite{IncorrectnessLogic}.

\subsection{Mixing types}

Goal: \emph{support mixed strict/nonstrict development}.

Like every active software community, Roblox developers share code
with one another constantly.  First- and third-party developers alike
frequently share entire software packages written in Luau.  To add to
this, many Roblox experiences are authored by a team.  It is therefore
crucial that we offer first-class support for mixing code written in
strict and nonstrict modes.

Some questions raised by mixed-mode types:
\begin{itemize}

\item How much feedback can we offer for a nonstrict script that is
  importing strict-mode code?

\item In strict mode, how do we talk about values and types that are
  drawn from nonstrict code?

\item How can we combine the goals of strict and nonstrict types?

\item Can we have strict and non-strict mode infer the same types,
  only with different flagging?

\item Is strict-mode code sound when it relies on non-strict code,
  which has weaker invariants?

\item How can we avoid introducing function wrappers in higher-order code
  at the strict/nonstrict boundary?

\end{itemize}
\emph{Related work}: there has been work on interoperability between different type systems,
notably~\cite{LinkingTypes}, but there the overall goals of the systems were similar safety properties.
In our case, the two type systems have different goals.

\subsection{Type inference}

Goal: \emph{infer types to allow gradual adoption of type annotations.}

Since backward compatibility with existing code is important, we have
to provide types for code without explicit annotations. Moreover, we
want to make use of type-directed tools such as autocomplete, so we
cannot adopt the common strategy of treating all untyped variables as
having type $\ANY$. This leads us to type inference.

To make use of type-driven technologies for programs
without explicit type annotations, we use a type inference algorithm.
Since Luau includes System~F, type inference is undecidable~\cite{Boehm85},
but we can still make use of heuristics such as local type inference~\cite{LocalTypeInference}.

It remains to be seen if type inference can satisfy the goals of
strict and non-strict types. The current Luau system
infers different types in the two modes, which is unsatisfactory as it
makes changing mode a non-local breaking change. In addition,
non-strict inference is currently too imprecise to support
type-directed tools such as autocomplete.

Some questions raised by type inference:
\begin{itemize}

\item How many cases in strict mode cannot be inferred by the type inference system? Minimizing
  this kind of error is desirable, to make the type system as unobtrusive as possible.
\item Can something like the Rust traits system~\cite{RustBook} or Haskell classes~\cite{TypeClasses} be used to provide types for overloaded operators, without hopelessly confusing learners?
\item Type inference currently infers monotypes for unannotated
  functions, in contrast to QuickLook~\cite{QuickLook}, which can infer generic types.
  Will this be good enough for idiomatic Luau scripts?
\item Can type inference be used to infer the same types in strict and nonstrict mode, to ease migrating between modes, with the only difference being error reporting?
\end{itemize}
\emph{Related work}: there is a large body of work on type inference, largely summarized in~\cite{TAPL}.

\section{Conclusions}

In this paper, we have presented some of the goals of the Luau type
system, and how they map to the needs of the Roblox creator
community. We have also explored how these goals differ from traditional
type systems, where it is necessary to accommodate the unique needs of
the Roblox platform. We have sketched what a solution might look like;
all that remains is to draw the owl~\cite{HowToDrawAnOwl}.

\bibliographystyle{ACM-Reference-Format} \bibliography{bibliography}


\begin{thebibliography}{23}


\ifx \showCODEN    \undefined \def \showCODEN     #1{\unskip}     \fi
\ifx \showDOI      \undefined \def \showDOI       #1{#1}\fi
\ifx \showISBNx    \undefined \def \showISBNx     #1{\unskip}     \fi
\ifx \showISBNxiii \undefined \def \showISBNxiii  #1{\unskip}     \fi
\ifx \showISSN     \undefined \def \showISSN      #1{\unskip}     \fi
\ifx \showLCCN     \undefined \def \showLCCN      #1{\unskip}     \fi
\ifx \shownote     \undefined \def \shownote      #1{#1}          \fi
\ifx \showarticletitle \undefined \def \showarticletitle #1{#1}   \fi
\ifx \showURL      \undefined \def \showURL       {\relax}        \fi
\providecommand\bibfield[2]{#2}
\providecommand\bibinfo[2]{#2}
\providecommand\natexlab[1]{#1}
\providecommand\showeprint[2][]{arXiv:#2}

\bibitem[\protect\citeauthoryear{Brady}{Brady}{2017}]%
        {TDDIdris}
\bibfield{author}{\bibinfo{person}{Edwin Brady}.}
  \bibinfo{year}{2017}\natexlab{}.
\newblock \bibinfo{booktitle}{\emph{Type-Driven Development with {Idris}}}.
\newblock \bibinfo{publisher}{Manning}.
\newblock
\showISBNx{9781617293023}


\bibitem[\protect\citeauthoryear{Findler and Felleisen}{Findler and
  Felleisen}{2002}]%
        {Contracts}
\bibfield{author}{\bibinfo{person}{Robert~B. Findler} {and}
  \bibinfo{person}{Matthias Felleisen}.} \bibinfo{year}{2002}\natexlab{}.
\newblock \showarticletitle{Contracts for Higher-order Functions}. In
  \bibinfo{booktitle}{\emph{Proc. Int. Conf. Functional Programming}}.
  \bibinfo{pages}{48--59}.
\newblock


\bibitem[\protect\citeauthoryear{Hall, Hammond, Peyton~Jones, and Wadler}{Hall
  et~al\mbox{.}}{1996}]%
        {TypeClasses}
\bibfield{author}{\bibinfo{person}{Cordelia~V. Hall}, \bibinfo{person}{Kevin
  Hammond}, \bibinfo{person}{Simon~L. Peyton~Jones}, {and}
  \bibinfo{person}{Philip~L. Wadler}.} \bibinfo{year}{1996}\natexlab{}.
\newblock \showarticletitle{Type Classes in Haskell}.
\newblock \bibinfo{journal}{\emph{ACM Trans. Program. Lang. Syst.}}
  \bibinfo{volume}{18}, \bibinfo{number}{2} (\bibinfo{year}{1996}),
  \bibinfo{pages}{109–138}.
\newblock


\bibitem[\protect\citeauthoryear{Heeren}{Heeren}{2005}]%
        {TopQuality}
\bibfield{author}{\bibinfo{person}{Bastiaan~J. Heeren}.}
  \bibinfo{year}{2005}\natexlab{}.
\newblock \emph{\bibinfo{title}{Top Quality Type Error Messages}}.
\newblock \bibinfo{thesistype}{Ph.D. Dissertation}. \bibinfo{school}{U.
  Utrecht}.
\newblock


\bibitem[\protect\citeauthoryear{Klabnik, Nichols, and the
  Rust~Community}{Klabnik et~al\mbox{.}}{2021}]%
        {RustBook}
\bibfield{author}{\bibinfo{person}{Steve Klabnik}, \bibinfo{person}{Carol
  Nichols}, {and} \bibinfo{person}{the Rust~Community}.}
  \bibinfo{year}{2021}\natexlab{}.
\newblock \bibinfo{title}{The Rust Programming Language}.
\newblock
\newblock
\urldef\tempurl%
\url{https://doc.rust-lang.org/book/}
\showURL{%
\tempurl}


\bibitem[\protect\citeauthoryear{Lindahl and Sagonas}{Lindahl and
  Sagonas}{2006}]%
        {SuccessTyping}
\bibfield{author}{\bibinfo{person}{Tobias Lindahl} {and}
  \bibinfo{person}{Konstantinos Sagonas}.} \bibinfo{year}{2006}\natexlab{}.
\newblock \showarticletitle{Practical Type Inference Based on Success Typings}.
  In \bibinfo{booktitle}{\emph{Proc. Int. Conf. Principles and Practice of
  Declarative Programming}}. \bibinfo{pages}{167–178}.
\newblock


\bibitem[\protect\citeauthoryear{Lua.org and {PUC}-Rio}{Lua.org and
  {PUC}-Rio}{2021}]%
        {Lua}
\bibfield{author}{\bibinfo{person}{Lua.org} {and} \bibinfo{person}{{PUC}-Rio}.}
  \bibinfo{year}{2021}\natexlab{}.
\newblock \bibinfo{title}{The {Lua} Programming Language}.
\newblock
\newblock
\urldef\tempurl%
\url{https://lua.org}
\showURL{%
\tempurl}


\bibitem[\protect\citeauthoryear{McAdam}{McAdam}{2002}]%
        {RepairingTypeErrors}
\bibfield{author}{\bibinfo{person}{Bruce~J. McAdam}.}
  \bibinfo{year}{2002}\natexlab{}.
\newblock \emph{\bibinfo{title}{Repairing Type Errors in Functional Programs}}.
\newblock \bibinfo{thesistype}{Ph.D. Dissertation}. \bibinfo{school}{U.
  Edinburgh}.
\newblock


\bibitem[\protect\citeauthoryear{Meme}{Meme}{2010}]%
        {HowToDrawAnOwl}
\bibfield{author}{\bibinfo{person}{Know~Your Meme}.}
  \bibinfo{year}{2010}\natexlab{}.
\newblock \bibinfo{title}{How To Draw An Owl}.
\newblock
\newblock
\urldef\tempurl%
\url{https://knowyourmeme.com/memes/how-to-draw-an-owl}
\showURL{%
\tempurl}


\bibitem[\protect\citeauthoryear{O'Hearn}{O'Hearn}{2020}]%
        {IncorrectnessLogic}
\bibfield{author}{\bibinfo{person}{Peter~W. O'Hearn}.}
  \bibinfo{year}{2020}\natexlab{}.
\newblock \showarticletitle{Incorrectness Logic}. In
  \bibinfo{booktitle}{\emph{Proc. Symp. Principles of Programming Languages}}.
  Article \bibinfo{articleno}{10}, \bibinfo{numpages}{32}~pages.
\newblock


\bibitem[\protect\citeauthoryear{Omar, Voysey, Chugh, and Hammer}{Omar
  et~al\mbox{.}}{2019}]%
        {Hazel}
\bibfield{author}{\bibinfo{person}{Cyrus Omar}, \bibinfo{person}{Ian Voysey},
  \bibinfo{person}{Ravi Chugh}, {and} \bibinfo{person}{Matthew Hammer}.}
  \bibinfo{year}{2019}\natexlab{}.
\newblock \showarticletitle{Live Functional Programming with Typed Holes}. In
  \bibinfo{booktitle}{\emph{Proc. Symp. Principles of Programming Languages}}.
  \bibinfo{pages}{14:1--14:28}.
\newblock


\bibitem[\protect\citeauthoryear{Patterson and Ahmed}{Patterson and
  Ahmed}{2017}]%
        {LinkingTypes}
\bibfield{author}{\bibinfo{person}{Daniel Patterson} {and}
  \bibinfo{person}{Amal Ahmed}.} \bibinfo{year}{2017}\natexlab{}.
\newblock \showarticletitle{Linking Types for Multi-Language Software: Have
  Your Cake and Eat It Too}. In \bibinfo{booktitle}{\emph{Proc. Summit on
  Advances in Programming Languages}}.
\newblock


\bibitem[\protect\citeauthoryear{Pierce}{Pierce}{2002}]%
        {TAPL}
\bibfield{author}{\bibinfo{person}{Benjamin~C. Pierce}.}
  \bibinfo{year}{2002}\natexlab{}.
\newblock \bibinfo{booktitle}{\emph{Types and Programming Languages}}.
\newblock \bibinfo{publisher}{{MIT} Press}.
\newblock
\showISBNx{0-262-16209-1}


\bibitem[\protect\citeauthoryear{Pierce and Turner}{Pierce and Turner}{2000}]%
        {LocalTypeInference}
\bibfield{author}{\bibinfo{person}{Benjamin~C. Pierce} {and}
  \bibinfo{person}{David~N. Turner}.} \bibinfo{year}{2000}\natexlab{}.
\newblock \showarticletitle{Local Type Inference}.
\newblock \bibinfo{journal}{\emph{ACM Trans. Program. Lang. Syst.}}
  \bibinfo{volume}{22}, \bibinfo{number}{1} (\bibinfo{year}{2000}),
  \bibinfo{pages}{1–44}.
\newblock


\bibitem[\protect\citeauthoryear{polymorphic type inference~is
  undecidable}{polymorphic type inference~is undecidable}{1985}]%
        {Boehm85}
\bibfield{author}{\bibinfo{person}{Partial polymorphic type inference~is
  undecidable}.} \bibinfo{year}{1985}\natexlab{}.
\newblock \showarticletitle{Hans-J. Boehm}. In \bibinfo{booktitle}{\emph{Proc.
  Symp. Foundations of Computer Science}}. \bibinfo{pages}{339--345}.
\newblock


\bibitem[\protect\citeauthoryear{Roblox}{Roblox}{2020}]%
        {RobloxDevelopers}
\bibfield{author}{\bibinfo{person}{Roblox}.} \bibinfo{year}{2020}\natexlab{}.
\newblock \bibinfo{title}{Roblox Developers Expected to Earn Over \$250 Million
  in 2020; Platform Now Has Over 150 Million Monthly Active Users}.
\newblock
\newblock
\urldef\tempurl%
\url{https://corp.roblox.com/2020/07/roblox-developers-expected-earn-250-million-2020-platform-now-150-million-monthly-active-users/}
\showURL{%
\tempurl}


\bibitem[\protect\citeauthoryear{Roblox}{Roblox}{2021a}]%
        {Luau}
\bibfield{author}{\bibinfo{person}{Roblox}.} \bibinfo{year}{2021}\natexlab{a}.
\newblock \bibinfo{title}{The {Luau} Programming Language}.
\newblock
\newblock
\urldef\tempurl%
\url{https://luau-lang.org}
\showURL{%
\tempurl}


\bibitem[\protect\citeauthoryear{Roblox}{Roblox}{2021b}]%
        {AllEducators}
\bibfield{author}{\bibinfo{person}{Roblox}.} \bibinfo{year}{2021}\natexlab{b}.
\newblock \bibinfo{title}{Roblox Education: All Educators}.
\newblock
\newblock
\urldef\tempurl%
\url{https://education.roblox.com/en-us/educators}
\showURL{%
\tempurl}


\bibitem[\protect\citeauthoryear{Roblox}{Roblox}{2021c}]%
        {Roblox}
\bibfield{author}{\bibinfo{person}{Roblox}.} \bibinfo{year}{2021}\natexlab{c}.
\newblock \bibinfo{title}{What is {Roblox}}.
\newblock
\newblock
\urldef\tempurl%
\url{https://corp.roblox.com}
\showURL{%
\tempurl}


\bibitem[\protect\citeauthoryear{Serrano, Hage, Peyton~Jones, and
  Vytiniotis}{Serrano et~al\mbox{.}}{2020}]%
        {QuickLook}
\bibfield{author}{\bibinfo{person}{Alejandro Serrano},
  \bibinfo{person}{Jurriaan Hage}, \bibinfo{person}{Simon Peyton~Jones}, {and}
  \bibinfo{person}{Dimitrios Vytiniotis}.} \bibinfo{year}{2020}\natexlab{}.
\newblock \showarticletitle{A quick look at impredicativity}. In
  \bibinfo{booktitle}{\emph{Proc. Int. Conf. Functional Programming}}.
\newblock


\bibitem[\protect\citeauthoryear{Siek and Taha}{Siek and Taha}{2006}]%
        {GradualTyping}
\bibfield{author}{\bibinfo{person}{Jeremy~G. Siek} {and} \bibinfo{person}{Walid
  Taha}.} \bibinfo{year}{2006}\natexlab{}.
\newblock \showarticletitle{Gradual Typing for Functional Languages}. In
  \bibinfo{booktitle}{\emph{Proc. Scheme and Functional Programming Workshop}}.
  \bibinfo{pages}{81--92}.
\newblock


\bibitem[\protect\citeauthoryear{Tobin-Hochstadt, Felleisen, Findler, Flatt,
  Greenman, Kent, St-Amour, Strickland, and Takikawa}{Tobin-Hochstadt
  et~al\mbox{.}}{2017}]%
        {MigratoryTyping}
\bibfield{author}{\bibinfo{person}{Sam Tobin-Hochstadt},
  \bibinfo{person}{Matthias Felleisen}, \bibinfo{person}{Robert~Bruce Findler},
  \bibinfo{person}{Matthew Flatt}, \bibinfo{person}{Ben Greenman},
  \bibinfo{person}{Andrew~M. Kent}, \bibinfo{person}{Vincent St-Amour},
  \bibinfo{person}{T.~Stephen Strickland}, {and} \bibinfo{person}{Asumu
  Takikawa}.} \bibinfo{year}{2017}\natexlab{}.
\newblock \showarticletitle{Migratory Typing: Ten Years Later}. In
  \bibinfo{booktitle}{\emph{Proc. Summit on Advances in Programming
  Languages}}.
\newblock


\bibitem[\protect\citeauthoryear{Wadler and Findler}{Wadler and
  Findler}{2009}]%
        {WellTyped}
\bibfield{author}{\bibinfo{person}{Philip Wadler} {and}
  \bibinfo{person}{Robert~B. Findler}.} \bibinfo{year}{2009}\natexlab{}.
\newblock \showarticletitle{Well-typed Programs Can’t be Blamed}. In
  \bibinfo{booktitle}{\emph{Proc. European Symp. Programming}}.
  \bibinfo{pages}{1--16}.
\newblock


\end{thebibliography}

\end{document}